\documentstyle[12pt,epsfig]{article}

\begin{document}
\begin{flushright}
hep-ph/0607335 \\
RAL-TR-2006-012 \\
30 July 2006 \\
\end{flushright}
\vspace{2 mm}
\begin{center}
{\Large
Real Invariant Matrices 
and
Flavour-Symmetric Mixing Variables
with Emphasis on Neutrino Oscillations}
\end{center}
\vspace{3mm}
\begin{center}
{P. F. Harrison\\
Department of Physics, University of Warwick,\\
Coventry, CV4 7AL. UK \footnotemark[1]}
\end{center}
\begin{center}
{W. G. Scott\\
CCLRC Rutherford Appleton Laboratory,\\
Chilton, Didcot, Oxon OX11 0QX. UK \footnotemark[2]}
\end{center}
\begin{center}
{T. J. Weiler \\
Department of Physics and Astronomy,\\
Vanderbilt University, Nashville, TN37235, USA. \footnotemark[3]}
\end{center}
\vspace{3mm}
\begin{abstract}
\baselineskip 0.6cm
\noindent
In 
fermion mixing phenomenology,
the matrix of moduli squared, $P= (|U|^2)$,
is well-known to carry
essentially the same information as the complex mixing matrix $U$ itself,
but with the advantage of being 
phase-convention independent.
The matrix $K$ 
(analogous to the Jarlskog $CP$-invariant $J$) 
formed from the 
real parts of the mixing matrix  
``plaquette'' products 
is similarly invariant. 
In this paper,
the $P$ and $K$ matrices 
are shown to be entirely equivalent,
both being directly related 
(in the leptonic case) to
the observable, locally $L/E$-averaged
transition probabilities in neutrino oscillations.
We study an (over-)complete set of 
flavour-symmetric Jarlskog-invariant functions
of mass-matrix commutators, 
rewriting them simply as moment-transforms 
of such (real) invariant matrices.
\end{abstract}
\begin{center}
\end{center}
\footnotetext[1]{E-mail:p.f.harrison@warwick.ac.uk}
\footnotetext[2]{E-mail:w.g.scott@rl.ac.uk}
\footnotetext[3]{E-mail:tom.weiler@vanderbilt.edu}
\newpage
\baselineskip 0.6cm

\noindent {\bf 1. Introduction: The P matrix}
\vspace{2mm}

\noindent
The complex mixing matrix 
(CKM matrix \cite{ckm} or MNS matrix \cite{mns} respectively)
occupies pride of place
at the center of quark and lepton mixing phenomenology.
The mixing matrix connects mass-eigenstates
of different charges, for example, 
charged-leptons to neutrinos, 
through an array of complex mixing amplitudes:
\begin{equation}
U= \left( \matrix{ U_{e1} & U_{e2} & U_{e3} \cr
            U_{\mu 1} & U_{\mu 2} & U_{\mu 3} \cr
            U_{\tau 1} & U_{\tau 2} & U_{\tau 3} } \right), \label{defu}
\end{equation}
where our choice to display the mixing matrix 
for leptons here, rather than quarks,
anticipates our emphasis on neutrino oscillations below.
Of course,
the freedom to rephase rows and columns 
of the mixing matrix 
means that the mixing-matrix elements
(in this case the $U_{l \nu}$ where $l=e, \mu, \tau$ and $\nu=1, 2, 3$)
are phase-convention dependent.

While no one would deny the importance
of amplitudes and
the mixing-matrix concept,
it has long been appreciated \cite{hamz} that
observables 
(including the magnitudes of $CP$-violating asymmetries
\footnote{Of course the information on the {\em sign} 
of the relevant Jarlskogian $J$ (for quarks or leptons) 
is lost in taking the moduli,
with a consequent overall sign ambiguity affecting
all associated $CP$-asymmetries 
(for quarks or leptons respectively).
In principal, however,
this single qbit of information (per sector)
could be carried in an {\it ad hoc} fashion
by giving a sign to the $P$-matrix itself,
which will anyway be taken as positive in any other context. 
We stress that we expect the considerations of this paper
to be valid for both Dirac and Majorana neutrinos,
provided we restrict attention 
to flavour oscillation phenomena,
neglecting neutrinoless double beta-decay observables.   
} 
see below) 
are always ultimately expressible in terms
of moduli-squared of mixing elements,
whereby we often find it convenient
to work instead with the corresponding 
matrix of moduli-squared,
$P = (|U|^2) = U \circ U^*$ 
(where the ``$\circ$'' denotes the simple 
entrywise (Schur) product of two matrices):
\begin{equation}
P \; = \; (|U|^2) \; = \; U \circ U^* \; = \; 
\left( \matrix{ |U_{e1}|^2 & |U_{e2}|^2 & |U_{e3}|^2 \cr
                |U_{\mu 1}|^2 & |U_{\mu 2}|^2 & |U_{\mu 3}|^2 \cr
                |U_{\tau 1}|^2 & |U_{\tau 2}|^2 & |U_{\tau 3}|^2 } \right). \label{pmat}
\end{equation} 
The $P$-matrix represents rather 
the probabilistic content of the various states
(again, in this case, the charged-letpton mass-eigenstates
versus the neutrino mass-eigenstates)
and has the advantage of being independent
of any choice of phase convention.

The $P$-matrix for leptons (Eq.~\ref{pmat})
features naturally
in neutrino oscillation phenomenology.
We have, eg.\ that $P \, P^T$ 
(where $T$ denotes the matrix transpose)  
represents directly \cite{ppT} the 
flavour-to-flavour transition/oscillation-probability matrix ${\cal P}(L/E)$
\begin{equation}
< \! \! {\cal P} \! \! >_{\infty} \; \; := \;
\lim_{L/E \rightarrow \infty}
< \! \! {\cal P}(L/E) \! \!> 
\hspace{4mm}  
\; = \; P P^T \label{ppt}
\end{equation}
for neutrino oscillations
in the ``asymptotic'' domain ($\infty$),
ie.\ when $L/E >> (\Delta m_{ij}^2)^{-1}$ 
for all neutrino mass-squared differences $\Delta m_{ij}^2$, $i,j=1-3$,
with local $L/E$-averaging
($L$ the propagation length and $E$ the neutrino energy).
The corresponding result
for the ``intermediate'' domain ($\Join$),
when
$(\Delta m_{12}^2)^{-1} >> L/E >> (\Delta m_{23}^2)^{-1},(\Delta m_{31}^2)^{-1}$,
again averaging over unresolved oscillations within the domain,
is also expressible entirely in terms of the $P$-matrix (see below). 
The transition/oscillation probability matrix ${\cal P}(L/E)$
gives disappearance probabilities on the diagonal 
and appearance probabilities off the diagonal, 
and (given the sign of $J$)  
its full $L/E$-dependent form, ie.\ with no $L/E$-averaging, 
is also expressible in terms of the $P$-matrix, see Eq.~\ref{probp} below.

We begin by noting that 
any two rows or columns of the $P$-matrix
are sufficient to calculate the 
magnitude of the Jarlskog $CP$-violation parameter \cite{jcp} using, respectively:
\begin{eqnarray}
4J^2 & = & (P_{l 1} \, P_{l' 1} \, + \, P_{l 2} \, P_{l' 2} \, + \, P_{l 3} \, P_{l' 3} \, )^2
 \, - \, 2 \, ( P_{l 1}^2 \, P_{l' 1}^2 \, + \, P_{l 2}^2 \, P_{l' 2}^2 \, + \, P_{l 3}^2 \, P_{l' 3}^2 \, ) 
                                                                     \label{ptojr} \\
4J^2 & = & (P_{e \nu} P_{e \nu'}+P_{\mu \nu} P_{\mu \nu'}+P_{\tau \nu} P_{\tau \nu'} \,)^2
        - \, 2 \, (P_{e \nu}^2 P_{e \nu'}^2+P_{\mu \nu}^2 P_{\mu \nu'}^2+P_{\tau \nu}^2 P_{\tau \nu'}^2)
                                                                     \label{ptojc}
\end{eqnarray}
where $l \neq l'$ and $\nu \neq \nu'$. 
Symmetrising over, eg.\  rows,
and exploiting the Schur product
and the $3 \times 3$ identity matrix $I$ together with its binary complement~$\bar{I}$
\begin{equation}
I = \left( \matrix{1 & 0 & 0 \cr 0 & 1 & 0 \cr 0 & 0 & 1} \right) \hspace{1.5cm}
\bar{I} = \left( \matrix{0 & 1 & 1 \cr 1 & 0 & 1 \cr 1 & 1 & 0} \right),
\end{equation}
we obtain a relatively simple matrix expression for $J^2$:
\begin{equation}
J^2= \frac{1}{24}{\rm Tr}  [ (\bar{I} \circ (P P^T))(\bar{I} \circ (P P^T))
                           -2\bar{I}(\bar{I} \circ ((P \circ P) (P \circ P)^T)) ]
\label{p4ov24} \end{equation}
which is quartic in $P$ and obviously $P \leftrightarrow P^T$ symmetric.
   
The full time-dependent flavour-to-flavour vacuum
transition/oscillation probability matrix ${\cal P}(L/E)$
is (`dimensionally') quadratic in $P$ and may now be written:
\begin{eqnarray}
{\cal P}(L/E) 
&  =  & I \circ \, ( \, P
\left( \matrix{1 & c_{12} & c_{31} \cr c_{12} & 1 & c_{23} \cr c_{31} & c_{23} & 1} \right) P^T \, )
\nonumber \\
 & + &   \bar{I} \circ ( \, P 
\left( \matrix{ 1-c_{12}+c_{23}-c_{31} \hspace{-5mm} &  0  &  0  \cr
                0  &\hspace{-2.5mm} 1-c_{12} -c_{23} +c_{31} \hspace{-2.5mm} &  0 \cr
                0  &  0  & \hspace{-5mm}1+c_{12}-c_{23} -c_{31} } 
        \right) P^T \, ) \nonumber \\  &  & \hspace{3.5cm} + \hspace{3mm} 
2 \,J \,  \left( \matrix{ 0 & 1 & -1 \cr -1 & 0 & 1 \cr 1 & -1 & 0 } \right)
                                          \hspace{1mm}  (s_{12}+s_{23}+s_{31}) \label{probp}
\end{eqnarray} 
where the first term gives 
the disappearance probabilities and the second and third terms
the $CP$-even and $CP$-odd parts of the appearance probabilities
(the latter contributing proportional to the `epsilon matrix' $\epsilon$ \cite{hsim}).
The time dependence 
in Eq.~\ref{probp}
enters only through
$c_{ij}:=\cos \Delta m^2_{ij}L/2E$ and
$s_{ij}:=\sin \Delta m^2_{ij}L/2E$.
The asymptotic limit 
is obtained setting 
$< \! \! c_{ij} \! \!>_{\infty} \, = \, < \! \! s_{ij} \! \! >_{\infty} \, = \, 0$ ($i,j=1-3$)
when Eq.~\ref{probp} reduces to $PP^T$ as expected.
The corresponding probabilities in the intermediate domain
are obtained setting $< \! \! c_{12} \! \! >_{\Join} \, = \, 1, 
\; < \! \! s_{12} \! \! >_{\Join} \,= \, \; 
< \! \! c_{23} \! \! >_{\Join} \; = \; < \! \! s_{23} \! \! >_{\Join} \, =  \,
< \! \! c_{31} \! \! >_{\Join} \, = \, < \! \! s_{31} \! \! >_{\Join} \; = \; 0$.

We may remark that
average appearance probaility measurements in the intermediate domain
determine directly the $\nu_3$ column of the $P$-matrix,
via $< \! \! {\cal P}_{e \mu} \! \! >_{\Join} =2P_{e3}P_{\mu 3}$,
$< \! \! {\cal P}_{\mu \tau} \! \! >_{\Join} =2P_{\mu 3}P_{\tau 3}$ etc.
whereby $P_{e3}=[< \! \! {\cal P}_{e \mu} \! \! >_{\Join}
< \! \! {\cal P}_{e \tau} \! \! >_{\Join}/
                        < \! \! {\cal P}_{\mu \tau} \! \! >_{\Join}]^{\frac{1}{2}}$ etc.
(a zero in the $P$-matrix here 
leads to indeterminacies, in which case
we must use disappearance probabilities also:
$< \! \! {\cal P}_{\mu \mu} \! \! >_{\Join}
= 1-2P_{\mu3}+2P_{\mu3}^2$ $\Rightarrow$ 
$P_{\mu 3}=1/2 \pm [1/4-(1-< \! \! {\cal P}_{\mu \mu} \! \! >_{\Join})/2]^{\frac{1}{2}}$).
Disappearance probabilities in the asymptotic domain
then determine the remaing independent column:
$< \! \! {\cal P}_{e e} \! \! >_{\infty} \;
=P_{e1}^2+P_{e2}^2+P_{e3}^2$, 
$< \! \! {\cal P}_{\mu \mu} \! \! >_{\infty} \;
=P_{\mu 1}^2+P_{\mu 2}^2+P_{\mu 3}^2$ etc.
whereby $P_{e \, 1(2)}=(1-P_{e3})/2 \pm 
[(1-P_{e3})^2/4+P_{e3}(1-P_{e3})-(1-< \! \! {\cal P}_{e e} \! \! >_{\infty})/2]^{\frac{1}{2}}$ etc.\ 
(sign ambiguities correlate
with the interchange of entries 
between $P$-matrix columns).
Note that
a complete set of
average vacuum probability measurements,
in the intermediate and asymptotic domains together, 
in general over-determines the $P$-matrix
and so fixes the magnitude of 
$J$ (via Eq.~\ref{p4ov24}),
despite the fact that the $CP$-violating third term
in Eq.~\ref{probp} makes no contribution 
to averaged probabilities in either domain.

Clearly we must fore-go here
any detailed discussion of experimental data,
measurement errors, matter effects etc.
Indeed it will suffice, for the purposes of this paper,
simply to take as given, in the first instance, 
the succinct summary of the current oscillation data 
provided by the tri-bimaximal ansatz \cite{tbm1} \cite{tbm2}:
\begin{eqnarray}
U= \left( \matrix{ \sqrt{2/3} & 1/\sqrt{3} & 0 \cr
                   -1/\sqrt{6} & 1/\sqrt{3} & 1/\sqrt{2} \cr
                   -1/\sqrt{6} & 1/\sqrt{3} & -1/\sqrt{2} } \right)
\hspace{8mm} \Rightarrow \hspace{8mm}
P= \left( \matrix{ 2/3 & 1/3 & 0 \cr
                   1/6 & 1/3 & 1/2 \cr
                   1/6 & 1/3 & 1/2 } \right) \label{utoptbm}
\end{eqnarray}
where we display the mixing matrix 
and the corresponding $P$-matrix side-by-side.
From, eg.\ Eq.~\ref{ptojr}, applied to, eg.\
the first two rows of the $P$-matrix Eq.~\ref{utoptbm},
we recover:
\begin{equation}
J^2=\frac{((2/3)(1/6)+(1/3)(1/3))^2-2((2/3)^2(1/6)^2+(1/3)^2(1/3)^2)}{4}=0.
\end{equation}
Comparing the LHS and RHS of Eq.~\ref{utoptbm},
the $P$-matrix definitely provides
the more visually clear summary,
without the distracting arbitrary (unobservable) phases/signs.

It turns out that we shall also find it useful below
to refer, occasionally and for illustrative purposes,
to the older trimaximal ansatz \cite{trx}
(now ruled out by the data):
\begin{eqnarray}
U= \left( \matrix{ 1/\sqrt{3} & 1/\sqrt{3} & 1/\sqrt{3} \cr
                   \omega/\sqrt{3} & 1/\sqrt{3} & \bar{\omega}/\sqrt{3} \cr
                   \bar{\omega}/\sqrt{3} & 1/\sqrt{3} & \omega/\sqrt{3} } \right)
\hspace{8mm} \Rightarrow \hspace{8mm}
P= \left( \matrix{ 1/3 & 1/3 & 1/3 \cr
                   1/3 & 1/3 & 1/3 \cr
                   1/3 & 1/3 & 1/3 } \right) \label{utoptri}
\end{eqnarray}
($\omega = \exp (i 2 \pi/3)$ and $\bar{\omega}= \exp (-i 2 \pi/3)$
represent the complex cube-roots of unity). 
From Eq.~\ref{ptojr}/Eq.~\ref{ptojc}, 
applied to any two rows/columns of the $P$-matrix Eq.~\ref{utoptri}, we have:
\begin{equation}
J^2=\frac{((1/3)^2+(1/3)^2+(1/3)^2)^2-2((1/3)^4+(1/3)^4+(1/3)^4)}{4}=\frac{1}{108}
\end{equation}
corresponding, as expected, to maximal $CP$-violation
($J=\pm 1/6\sqrt{3}$ \cite{sixrt3}).

The extreme hierarchical form of the CKM matrix,
ie.~the smallness of the off-diagonal elements,
makes the $P$-matrix 
in the quark case much less useful
visually, eg.\ as a summry of data,
than it is in the leptonic case.
Since in this paper we shall be emphasising
the $P$-matrix and in particular
some useful matrices computable from it  
(see Sections 2-4) we shall henceforth focus 
essentially exclusively on the leptons.\\

\noindent {\bf 2. Plaquettes: The K matrix} 
\vspace{2mm}

\noindent
A mixing matrix ``plaquette'' \cite{bjpi}
is a 4-subset
of complex mixing matrix elements,
obtained by deleting one row and one column
of the mixing matrix.
Associating a given plaquette
with the element common to the deleted row and column,
there is clearly one plaquette 
`complementary to'
each mixing element.
The corresponding plaquette product may be defined:
$\Pi_{l\nu} :=
U_{l-1\; \nu-1} U_{l-1\; \nu+1}^* U_{l+1 \; \nu+1} U_{l+1 \; \nu-1}^*$
(generation indices
to be interpreted mod 3,
meaning that
when $l=$``$\tau$'' we have $l+1=$``$e$'' etc.).
The matrix of plaquette products (for leptons) may then be written:
\begin{eqnarray}
\Pi=\left( \matrix{ 
  U_{\tau 3} U_{\tau 2}^* U_{\mu 2} U_{\mu 3}^*
                      & U_{\tau 1} U_{\tau 3}^* U_{\mu 3} U_{\mu 1}^*
                                 & & U_{\tau 2} U_{\tau 1}^* U_{\mu 1} U_{\mu 2}^*  \cr
  U_{e 3} U_{e 2}^* U_{\tau 2} U_{\tau 3}^*
                      & U_{e 1} U_{e 3}^* U_{\tau 3} U_{\tau 1}^*
                                    & & U_{e 2} U_{e 1}^* U_{\tau 1} U_{\tau 2}^*   \cr
  U_{\mu 3} U_{\mu 2}^* U_{e 2} U_{e 3}^*
                      & U_{\mu 1} U_{\mu 3}^* U_{e 3} U_{e 1}^*
\label{Pi}                                    & & U_{\mu 2} U_{\mu 1}^* U_{e 1} U_{e 2}^*} \right).
\end{eqnarray}
Plaquette products are phase-convention independent
complex numbers, and, as is well-known \cite{cecj},
all have the same\footnote{
Very often \cite{cecj} alternating signs 
are found to enter here 
(so that the imaginary parts in Eq.~\ref{PiKJ} are given by $\pm J$).
However, with our cyclic definition (Eq.~\ref{Pi}),
there are no such alterning signs.}
imaginary part, 
equal to and defining the Jarlskog $CP$-violating parameter $J$ \cite{jcp}.
The real parts define the $K$-matrix ($\Pi_{l \nu} :=-K_{l\nu} + iJ$):
\begin{equation}
\Pi= -\left( \matrix{ K_{e1} & K_{e2} & K_{e3} \cr
            K_{\mu 1} & K_{\mu 2} & K_{\mu 3} \cr
            K_{\tau 1} & K_{\tau 2} & K_{\tau 3} } \right)
\hspace{2mm} + \hspace{2mm}
i \, \left( \matrix{ J & J & J \cr
                J & J & J \cr
                J & J & J } \right) \label{PiKJ}
\end{equation}
(where for our convenience we have introduced a minus sign
in the definition with respect to some previous work, 
$K_{l\nu} :=-K_{l' l''}^{\nu'\nu''} $\cite{kmat}).
The $K$-matrix \cite{weil}\cite{kmat}
may be viewed as the natural $CP$-conserving analogue 
of the Jarlskog variable $J$.

The $K$-matrix is an important oscillation observable.
The magnitnude of the Jarlskog $CP$ violation parameter may be obtained
directly \cite{weil} from the $K$-matrix by summing products in pairs of 
$K$-matrix elements within any row or column: 
\begin{eqnarray}
K_{l1}K_{l2}+K_{l2}K_{l3}+K_{l3}K_{l1}=J^2 \label{rowkj2} \\
K_{e \nu}K_{\mu \nu}+K_{\mu \nu} K_{\tau \nu} +K_{\tau \nu} K_{e \nu} =J^2. \label{colkj2}
\end{eqnarray}
Symmetrising over rows or columns we obtain respectively:
\begin{equation}
J^2= \frac{1}{6} {\rm Tr} \; [ \,\bar{I} K K^T \, ]
            = \frac{1}{6} {\rm Tr} \; [ \,\bar{I} K^T K \, ]
\end{equation} 
which is clearly very succinct 
and obviously $K \leftrightarrow K^T$ symmetric.

The full time-dependent flavour-to-flavour vacuum
transition/oscillation probability matrix 
is (`dimensionally') linear in $K$ and may be written:
\begin{eqnarray}
{\cal P}(L/E)  \;  = \; I \circ (I- 2 \bar{I} K
\left( \matrix{ 1-c_{23}  &  1-c_{23}  &  1-c_{23}  \cr
                1-c_{31}  &  1-c_{31} &  1-c_{31} \cr
                1-c_{12}  &  1-c_{12}  & 1-c_{12} }  \right)) \hspace{6.0cm}
\nonumber \\
\hspace{-0.2cm} +  \hspace{-0.0cm} I_+ \circ (2I_- K 
\left( \matrix{ 1-c_{23}  &  1-c_{23}  &  1-c_{23}  \cr
                1-c_{31}  &  1-c_{31} &  1-c_{31} \cr
                1-c_{12}  &  1-c_{12}  & 1-c_{12} } 
        \right)) + 
 I_- \circ (2I_+ K 
\left( \matrix{ 1-c_{23}  &  1-c_{23}  &  1-c_{23}  \cr
                1-c_{31}  &  1-c_{31} &  1-c_{31} \cr
                1-c_{12}  &  1-c_{12}  & 1-c_{12} } 
        \right)) \hspace{0.3cm} \nonumber \\  
 \hspace{2.5cm} + \hspace{3mm} 
2 \, J \, \left( \matrix{ 0 & 1 & -1 \cr -1 & 0 & 1 \cr 1 & -1 & 0 } \right)
                        \hspace{2mm}  (s_{12}+s_{23}+s_{31}) \hspace{1.0cm}
\label{probk}
\end{eqnarray} 
where we exploit the (cyclic) flavour raising and lowering operators:
\begin{equation}
I_+ = \left( \matrix{0 & 0 & 1 \cr 1 & 0 & 0 \cr 0 & 1 & 0} \right) \hspace{1.5cm}
I_- = \left( \matrix{0 & 1 & 0 \cr 0 & 0 & 1 \cr 1 & 0 & 0} \right)
\end{equation}
($\bar{I}=I_++I_-$ and $\epsilon=I_+-I_-$).
Note that, in Eq.~\ref{probk}, the second and third terms,
constituting the $CP$-even part of the appearance probabilities, 
are simply transposes one of the other.
The asymptotic limit 
may be written:
\begin{equation}
< \! \! {\cal P} \! \! >_{\infty} \; \;
= \; I +(I_+-I) \circ (2I_-KD)+(I_--I) \circ (2 I_+KD) \label{pkinf}
\end{equation} 
where $D$ is the so-called `democratic matrix' ($D=I+\bar{I}$),
and appearance probabilities 
are now given by the $K$-matrix row-sums: 
$< \! \! {\cal P}_{e \mu} \! \! >_{\infty} \, = \, 2(K_{\tau 1}+K_{\tau 2}+K_{\tau 3})$ etc.
In the intermediate domain,
appearance probabilities are 
given by:
$< \! \! {\cal P}_{e \mu} \! \! >_{\Join} \, = \, 2(K_{\tau 1}+K_{\tau 2})$ etc.\
so that the $\nu_3$ column is determined directly:
\mbox{ $K_{\tau 3} \, = \, (< \! \! {\cal P}_{e \mu} \! \! >_{\infty} \,
               - \, < \! \! {\cal P}_{e \mu} \! \! >_{\Join})/2$} etc.
and $J^2$ may be calculated from Eq.~\ref{colkj2}.
The remaining elements are given by:
$K_{\tau \, 1(2)} = \, < \! \! {\cal P}_{e \mu} \! \! >_{\Join}/4 
          \pm [< \! \! {\cal P}_{e \mu} \! \! >_{\Join}^2/16 
                + < \! \! {\cal P}_{e \mu} \! \! >_{\Join} K_{\tau 3}/2 - 4J^2]^{\frac{1}{2}}$,
where the sign ambiguity correlates with the interchange
of the $\nu_1$ and $\nu_2$ columns, as before.

The $K$-matrx is readily computable~\cite{weil} \cite{kmat} 
from the $P$-matrix
(again, indices to be interpreted mod 3) using:
\begin{equation}
K_{l\nu}=(P_{l\nu}
 -(P_{l-1\; \nu-1} \; P_{l+1 \; \nu+1}
                    +P_{l-1\; \nu+1} \; P_{l+1 \; \nu-1}))/2, \label{keqp}
\end{equation}
ie.\ just take the corresponding element of the $P$-matrix, 
subtract the permanent of its associated ``$P$ plaquette''
(defined analogously to a mixing matrix plaquette above),
and divide by an overall factor of two
(the permanent of a matrix 
is defined similarly to its determinant, but
with the alternating signs replaced by all positive signs).

Given instead the $K$-matrix
we can also always obtain the $P$-matrix,
in effect inverting Eq.~\ref{keqp}.
Specifically, to obtain the $P$ matrix element $P_{l \nu}$, 
we focus on the corresponding $K$ matrix element $K_{l \nu}$  
and especially on its complementary ``$K$-plaquette''
(again defined analogously to a mixing matrix plaquette).
We sum the products of adjacent elements of the $K$-plaquette 
giving $A_{l\nu}$ 
and add twice its permanent, $M_{l\nu}$:
\begin{eqnarray}
A_{l\nu}=(K_{l+1 \; \nu +1}+K_{l-1 \nu-1})(K_{l-1 \; \nu+1}+K_{l+1 \; \nu-1}) \\
M_{l\nu}=K_{l+1 \; \nu +1}K_{l-1\; \nu-1}+ K_{l-1 \; \nu +1}K_{l+1 \; \nu-1} \hspace{1.0cm}
\end{eqnarray}
($A_{l \nu}$ is the product of the sums of diagonally opposing elements of the $K$-plaquette 
while $M_{l \nu}$ is the sum of their products).
Finally we divide by the sum of all $K$-elements not in the $K$-plaquette 
(excluding $K_{l \nu}$ itself) and take the (positive) square root:
\footnote{Since the argument of this square root must always be positive
for a valid $K$-matrix, we could equally (c.f.\ footnote~1) carry the sign 
of the $CP$-violation adopting an ad-hoc sign convention for~$K$.
(For non-zero mixing, the sum of all $K$-matrix elements 
is always positive in our present convention.)}   
\begin{equation}
P_{l \; \nu}= \sqrt{(A_{l\nu}+2M_{l\nu})/(K_{l+1 \; \nu}
                                              +K_{l-1 \; \nu}+K_{l \; \nu+1} +K_{l \; \nu-1})}.
\end{equation}

To illustrate the above procedures 
we can take the tri-bimaximal ansatz as a convenient example,
and display the P matrix and the K matrix side-by-side.:
\begin{eqnarray}
P= \left( \matrix{ 2/3 & 1/3 & 0 \cr
                   1/6 & 1/3 & 1/2 \cr
                   1/6 & 1/3 & 1/2 } \right)
\hspace{8mm} \Leftrightarrow \hspace{8mm}
K= \left( \matrix{ 1/6 & 1/12 & -1/18 \cr
                    0 & 0 & 1/9 \cr
                    0 & 0 & 1/9 } \right). \label{ptbmk}
\end{eqnarray}
In computing $K$ from $P$ we have, eg.\ that
the $K_{e1}$ element is given by $P_{e1}$
less its complementary $P$-permanent, 
$P_{\mu 2} P_{\tau 3} + P_{\mu 3} P_{\tau 2}$, all divided by two:
\begin{equation}
K_{e 1} = (2/3-(1/3 \times 1/2+1/2 \times 1/3))/2 =(2/3-1/3)/2=1/6.
\end{equation}
In computing $P$ from $K$ we have, eg.\ that
for $P_{e 1}$ the $K$-permanent is zero, 
and there is only one non-zero adjacent-pair product
($P_{\mu 3} \times P_{\tau 3}$), whereby:
\begin{equation}
P_{e 1} = \sqrt{\frac{1/9 \times 1/9}{1/12-1/18}}
 =  \sqrt{\frac{1/9 \times 1/9}{1/36}}=\sqrt{\frac{36}{81}}=2/3
\end{equation}
with only two terms ($K_{e2}=1/12$, $K_{e3}=-1/18$) 
non-zero in the denominator sum. 
The Jarlskog parameter computed from the $K$-matrix Eq.~\ref{ptbmk},
eg.\ from the first row:
\begin{equation}
J^2=\left( \frac{1}{6}\right) \left( \frac{1}{12}\right) 
              +\left( \frac{1}{12} \right) \left( -\frac{1}{18} \right)
                    +\left( -\frac{1}{18} \right) \left( \frac{1}{6} \right) =0
\end{equation}
is seen to vanish, as expected for tribimaximal mixing.

Similarly (and for further illustration) 
we have for the trimaximal ansatz:
\begin{eqnarray}
P= \left( \matrix{ 1/3 & 1/3 & 1/3 \cr
                   1/3 & 1/3 & 1/3 \cr
                   1/3 & 1/3 & 1/3 } \right)
\hspace{8mm} \Leftrightarrow \hspace{8mm}
K= \left( \matrix{ 1/18 & 1/18 & 1/18 \cr
                   1/18 & 1/18 & 1/18 \cr
                   1/18 & 1/18 & 1/18 } \right) \label{ktri}
\end{eqnarray}
In computing $K$ from $P$ we have that, eg.\
$K_{e1}$ is given by:
\begin{equation}
K_{e 1} = (1/3-(1/3 \times 1/3+1/3 \times 1/3))/2 =(1/3-2/9)/2=1/18
\end{equation}
and in computing $P$ from $K$ we have, eg.\ for $P_{e 1}$:
\begin{equation}
P_{e 1} = \sqrt{\frac{4/(18 \times 18)+ 2 \times 2/(18 \times 18)}{4/18}}
 =  \sqrt{\frac{8/(18 \times 18)}{4/18}}=\sqrt{\frac{2}{18}}=1/3
\end{equation}
where clearly all elements of either matrix are equivalent in the trimaximal case.
From any row or column of the $K$-matrix, summing products of pairs of elements, 
the Jarlskog invariant is given by $J^2=3/18^2=1/(36 \times 3)$, 
ie. $J = \pm 1/(6 \sqrt{3})$ as expected.

So if we have the full $K$-matrix we can calculate the full $P$-matrix and vice-versa.
The $K$-matrix and the $P$-matrix are thus entirely equivalent,
both carrying complete information
about the mixing, including the $CP$ violation  (excepting only its sign). 
In Appendix~A we show that,
just as the $P$-matrix is determined by any $P$-plaquette,
the $K$-matrix is similarly determined
by any $K$-plaquette (up to a twofold ambiguity). \\

\noindent {\bf 3. Quadratic Commutator Invariants: The Q-matrix }
\vspace{2mm}

\noindent
The $P$, $K$ and $U$ matrices above 
all have rows and columns (respectively) 
labelled by the charged-lepton and neutrino mass-eigenstates.
For some applications, however,
flavour-symmetric mixing variables,
making no reference whatever to any particular basis,
could be considered more appropriate
(e.g.\ as input to 
the proposed ``extremisation'' programme \cite{ext1}).
The prototype for
such 
variables 
is undoubtedly the Jarlskog $CP$-violating invariant itself \cite{jcp},
written here for the leptonic case 
in terms of the commutator 
$C := -i[L,N]$ of the charged-lepton ($L$) 
and neutrino ($N$) mass matrices:
\begin{eqnarray}
\hspace{0.4cm}
{\rm Tr} \; C^3/3 \, = \;
-2 \; \; {\rm Det} \; {\rm diag} (\Delta_l) \; \; {\rm Det} \; {\rm diag} (\Delta_{\nu}) 
\; J \; \; := \; -2 \, L_{\Delta} \, N_{\Delta} \; J \hspace{0.9cm} \label{detc1} \\
\Delta_l^T=(m_{\mu}-m_{\tau},\, m_{\tau}-m_{e}, \, m_{e}-m_{\mu}), 
\hspace{0.5cm}
\Delta_{\nu}^T=(m_{2}-m_{3},\, m_{3}-m_{1}, \, m_{1}-m_{2}) \label{detc2}
\end{eqnarray} 
with $m_e$, $m_{\mu}$, $m_{\tau}$ the charged-lepton masses
and $m_1$, $m_2$, $m_3$ the neutrino masses. 
%
%
In this section and the final section respectively,
we shall examine, systematically, generalisations 
of the Jarlskog invariant,
to a range of quadratic and cubic commutator traces,
which turn out to be intimately related 
to the $K$ and $P$ matrices (Sections~1-2).

We begin by considering
quadratic commutator invariants \cite{bran} defined 
($ C_{m n} :=-i[L^m,N^n]$ \cite{bran}) as follows \cite{ext1}:
\begin{equation}
Q \; = \; - \frac{1}{2}
          \left( \matrix { {\rm Tr} \, C_{11}^2  & {\rm Tr} \, C_{11}C_{12} & {\rm Tr} \, C_{12}^2  \cr
                 {\rm Tr} \, C_{11}C_{21} &  {\rm Tr} \, C_{11}C_{22} & {\rm Tr} \, C_{12}C_{22}  \cr
                 {\rm Tr} \, C_{21}^2 & {\rm Tr} \, C_{21}C_{22} &  {\rm Tr} \, C_{22}^2 } \right),
\label{qmat}
\end{equation}
each trace clearly flavour-symmetric and vanishing in the case of zero mixing.
Row and column labels now correspond to successive powers 
of charged-lepton and neutrino mass matrices respectively. 
Note that powers higher than $L^2$ or $N^2$ need not be considered, since,
eg.\ $L^3$ can be always re-expressed in terms of $L^2$, $L$ and $L^0=I$,
by virtue of the characteristic equation.
Furthermore ${\rm Tr} \, [L^2,N][L,N^2]= {\rm Tr} \, [L,N][L^2,N^2]$,
so that there are indeed just nine such invariants.

One may then show,
eg.\ by use of the flavour projection operator technique \cite{proj}, 
that the 
$Q_{nm}$ ($n$,$m$ $=$ $1,2,3$)
are just suitably-defined
double-moments of the $K$-matrix:
\begin{equation}
Q_{nm}=\Delta_l^T \; {\rm diag}(\Delta_l) \;
                     ({\rm diag} \, \Sigma_l)^{n-1}                                    
                      \; K \;   ({\rm diag} \, \Sigma_{\nu})^{m-1} \;
                          {\rm diag}(\Delta_{\nu}) \; \Delta_{\nu} \label{qnmk} \hspace{14mm}
\end{equation}
with 
mass-sum vectors $\Sigma_l$, $\Sigma_{\nu}$
(cf.~the mass-difference vectors $\Delta_l$, $\Delta_{\nu}$ above) given by:
\begin{eqnarray}
\Sigma_l=(m_{\mu}+m_{\tau},\, m_{\tau}+m_{e}, \, m_{e}+m_{\mu}) \hspace{0.5cm}
\Sigma_{\nu}=(m_{2}+m_{3},\, m_{3}+m_{1}, \, m_{1}+m_{2}).
\end{eqnarray}
Eq.~\ref{qnmk} (taken together with the definition Eq.~\ref{qmat})
is the manifest analogue of Eq,~\ref{detc1} for the $CP$-conserving case.
While the $Q$-matrix is then simply a linear transform of the $K$-matrix,
with respect to appropriate lepton and neutrino Vandermonde matrices: 
\begin{eqnarray}
Q
 & = &  
   \left( \matrix{ 1 & 1 & 1 \cr
                          (m_{\mu}+m_{\tau}) & (m_{\tau}+m_e) & (m_e+m_{\mu}) \cr
                          (m_{\mu}+m_{\tau})^2 & (m_{\tau}+m_e)^2 & (m_e+m_{\mu})^2 } \right)
  ({\rm diag } \, \Delta_l)^2  \nonumber  \\ 
&  & \; K \; ( {\rm diag} \, \Delta_{\nu})^2 
                   \left( \matrix{ 1 & 1 & 1 \cr
                          (m_2+m_3) & (m_3+m_1) & (m_1+m_2) \cr
                          (m_2+m_3)^2 & (m_3+m_1)^2 & (m_1+m_2)^2 } \right)^T ,
  \hspace{5mm}
\end{eqnarray}
the $K$-matrix is clearly simply the corresponding inverse transform of the $Q$-matrix: 
\begin{eqnarray}
K
 & = &  \frac{({\rm diag } \Delta_l)^{-1}}{L_{\Delta}} 
   \left( \matrix{ (m_e+m_{\mu})(m_{\tau}+m_e) & -(2m_e+m_{\mu}+m_{\tau}) & 1 \cr
                          (m_{\mu}+m_{\tau})(m_e+m_{\mu}) & -(2m_{\mu}+m_{\tau}+m_e) & 1 \cr
                          (m_{\tau}+m_e)(m_{\mu}+m_{\tau}) & -(2m_{\tau}+m_e+m_{\mu}) & 1 } \right)
 \nonumber  \\
& Q & \left( \matrix{ (m_1+m_2)(m_3+m_1) & -(2m_1+m_2+m_3) & 1 \cr
                          (m_2+m_3)(m_1+m_2) & -(2m_2+m_3+m_1) & 1 \cr
                          (m_3+m_1)(m_2+m_3) & -(2m_3+m_1+m_2) & 1 } \right)^T
\frac{( {\rm diag} \Delta_{\nu})^{-1} }{N_{\Delta}} \hspace{5mm}
\end{eqnarray}
so that (for known masses) the $Q$-matrix and the $K$-matrix are entirely equivalent.

Of course $K$-elements are not all independent;
products of pairs of elements in any row or column sum to $J^2$ \cite{weil} \cite{kmat}.
Corresponding relations result for the $Q$-matrix:
\begin{eqnarray}
[\Sigma^2_{l-1}\Sigma^2_{l+1}(Q_{11}Q_{13}-Q_{12}^2)
 +\Sigma_{l-1}\Sigma_{l+1}(\Sigma_{l-1}+\Sigma_{l+1})(2Q_{12}Q_{22}-Q_{11}Q_{23}-Q_{13}Q_{21}) \nonumber \\
 +(\Sigma_{l-1}+\Sigma_{l+1})(2Q_{22}Q_{32}-Q_{21}Q_{33}-Q_{23}Q_{31})
                                                      +(Q_{31}Q_{33}-Q_{32}^2) \nonumber \\
         +\Sigma_{l-1}\Sigma_{l+1}(Q_{11}Q_{33}+Q_{13}Q_{31}-2Q_{12}Q_{32}
                         +4Q_{21}Q_{23}-4Q_{22}^2) \nonumber \\
        +\Delta_l^2(Q_{21}Q_{23}-Q_{22}^2)]/
         (\Delta_l \, L_{\Delta} 
                      N_{\Delta} 
)^2 = J^2
\end{eqnarray}
\begin{eqnarray}
[\Sigma^2_{\nu-1}\Sigma^2_{\nu+1}(Q_{11}Q_{31}-Q_{21}^2)
 +\Sigma_{\nu-1}\Sigma_{\nu+1}(\Sigma_{\nu-1}+\Sigma_{\nu+1})
                                  (2Q_{21}Q_{22}-Q_{11}Q_{32}-Q_{31}Q_{12}) \nonumber \\
 +(\Sigma_{\nu-1}+\Sigma_{\nu+1})(2Q_{22}Q_{23}-Q_{12}Q_{33}-Q_{32}Q_{13})
                                                      +(Q_{13}Q_{33}-Q_{23}^2) \nonumber \\
         +\Sigma_{\nu-1}\Sigma_{\nu+1}(Q_{11}Q_{33}+Q_{13}Q_{31}-2Q_{21}Q_{23}
                         +4Q_{12}Q_{32}-4Q_{22}^2) \nonumber \\
        +\Delta_\nu^2(Q_{12}Q_{32}-Q_{22}^2)]/
         (\Delta_\nu \, L_{\Delta} 
                        N_{\Delta} 
)^2 = J^2
\end{eqnarray}
for $l$ $=$ $e$, $\mu$, $\tau$, and $\nu$ $=$ $1$, $2$, $3$, respectively
(where for eg.\ $l=e$, $\Sigma_{l+1}=\Sigma_{\mu}=m_{\tau}+m_{e}$, 
$\Sigma_{l-1}= \Sigma_{\tau}=m_e+m_{\mu}$, $\Delta_l=\Delta_e=m_{\mu}-m_{\tau}$ etc.).
Equating such expressions among themselves yields relations between $Q$-matrix elements,
and in particular flavour-symmetric relations may be obtained,
eg.\ by summing over all $l$ and all $\nu$ respectively:
\begin{eqnarray}
N_{\Delta}^2[L_{P4}^2(Q_{11}Q_{13}-Q_{12}^2)
                       +2L_{P4}L_{P3}(2Q_{12}Q_{22}-Q_{11}Q_{23}-Q_{13}Q_{21}) \nonumber \\
 +2L_{P3}L_{P2}(2Q_{22}Q_{32}-Q_{21}Q_{33}-Q_{23}Q_{31})+L_{P2}^2(Q_{31}Q_{33}-Q_{32}^2) \nonumber \\
                        +L_{P6}(Q_{11}Q_{33}+Q_{13}Q_{31}-2Q_{12}Q_{32}
                         +4Q_{21}Q_{23}-4Q_{22}^2)+3L_{\Delta}^2Q_{21}Q_{23}] \nonumber \\
=L_{\Delta}^2[N_{P4}^2(Q_{11}Q_{31}-Q_{21}^2)
                         +2N_{P4}N_{P3}(2Q_{21}Q_{22}-Q_{11}Q_{32}-Q_{31}Q_{12}) \nonumber \\
      +2N_{P3}N_{P2}(2Q_{22}Q_{23}-Q_{12}Q_{33}-Q_{32}Q_{13})+N_{P2}^2(Q_{13}Q_{33}-Q_{23}^2) \nonumber \\
                        +N_{P6}(Q_{11}Q_{33}+Q_{13}Q_{31}-2Q_{21}Q_{23}
                         +4Q_{12}Q_{32}-4Q_{22}^2)+3N_{\Delta}^2Q_{12}Q_{32}] \label{linj}
\end{eqnarray}
(basically the relation Tr $\bar{I}KK^T$ = Tr $\bar{I} K^T K$).
Further flavour-symmetric relations could clearly be obtaned, 
eg.\ by equating the sums over products in pairs etc., 
but such higher-power relations 
would not seem to have much practical value here.  
Clearly only five relations among $Q$-matrix elements 
can be functionally independent in total.


The supplementary polynomials used in Eq.~\ref{linj} above are defined 
in terms of the traces of powers of mass matrices: 
$L_1 \; := \; {\rm Tr} \; L$, $L_2 \; := \; {\rm Tr} \; L^2$,
$N_1 \; := \; {\rm Tr} \; N$ etc.~\cite{ext1}.
For, eg.~the charged-leptons, we have:
\begin{eqnarray}
L_{\Sigma} & := & (L_1^3 - L_3)/3 =(m_e+m_{\mu})(m_{\mu}+m_{\tau})(m_{\tau}+m_e) \\
L_{P2} & := & (3L_2 - L_1^2)/2 = m_e^2+m_{\mu}^2+m_{\tau}^2-m_em_{\mu}-m_{\mu}m_{\tau}-m_{\tau}m_e \\ 
L_{P3} & := & (3L_3 - L_2L_1)/2 =m_e^3+m_{\mu}^3+m_{\tau}^3 \nonumber \\        
  & &     \hspace{-3mm} -(m_e^2m_{\mu}+m_{\mu}^2m_e)/2
                 -(m_{\mu}^2m_{\tau}+m_{\tau}^2m_{\mu})/2-(m_{\tau}^2m_e+m_e^2m_{\tau})/2 \\
L_{P4} & := & L_{P2}^2+2(L_1L_3 - L_2^2) = m_e^4+m_{\mu}^4+m_{\tau}^4
                              -m_e^2m_{\mu}^2-m_{\mu}^2m_{\tau}^2-m_{\tau}^2m_e^2 \\
L_{\Delta}^2 & := & 4(L_{P4} \, L_{P2} \; - \; L_{P3}^2)/3
                                   = (m_e-m_{\mu})^2(m_{\mu}-m_{\tau})^2(m_{\tau}-m_e)^2\\
L_{P6} & := & L_{P3}^2 - L_{\Delta}^2/4 \nonumber
       = (m_e^3+m_{\mu}^3+m_{\tau}^3-m_e^2m_{\mu}-m_{\mu}^2m_{\tau}-m_{\tau}^2m_e) \nonumber \\
      &     & \hspace{2.5cm} \times (m_e^3+m_{\mu}^3+m_{\tau}^3-m_em_{\mu}^2-m_{\mu}m_{\tau}^2-m_{\tau}m_e^2)
\end{eqnarray}
with analogous relations for the neutrino polynomials in terms of $N_1$, $N_2$, $N_3$. \\

\noindent {\bf 4. Cubic Commutator Invariants: P and R Matrix Moments} 
\vspace{2mm}

\noindent Finally,
we consider all cubic commutator trace invariants,
completely generalising the original Jarlskog $CP$-invariant
(as for the mass-matrices themselves, 
powers of commutators higher than cubic
can aways be reduced via the characteristic equation).

Taking into account the cyclic property of the trace
we may classify the cubic commutator invariants,
according to the number of repeated commutators entering.
For example, there are four such invariants with all three comutators identical:
\begin{eqnarray}
-i{\rm Tr} \; C_{11}^3 & = & 6iL_{\Delta} N_{\Delta} J \label{c3eq1} \\
-i{\rm Tr} \; C_{12}^3 & = & 6iN_{\Sigma} L_{\Delta} N_{\Delta} J \label{c3eq2} \\
-i{\rm Tr} \; C_{21}^3 & = & 6iL_{\Sigma} L_{\Delta} N_{\Delta} J \label{c3eq3} \\
-i{\rm Tr} \; C_{22}^3 & = & 6iL_{\Sigma} N_{\Sigma} L_{\Delta} N_{\Delta} J \label{c3eq4}
\end{eqnarray}
all clearly proportional to $J$, 
and carrying no information on mixing angles beside 
the value of $J$ itself, assuming the masses are known.

Then one should consider the twelve cubic commutator invariants
with two commuatators identical and one different: 
\begin{eqnarray}
-i{\rm Tr} \; C_{11}^2C_{12} & = & 4iN_1 L_{\Delta} N_{\Delta} J \label{c32eq1} \\
-i{\rm Tr} \; C_{11}^2C_{21} & = & 4iL_1 L_{\Delta} N_{\Delta} J \label{c32eq2} \\
-i{\rm Tr} \; C_{12}^2C_{11} & = & 2iN_{P2} L_{\Delta} N_{\Delta} J \label{c32eq3} \\
-i{\rm Tr} \; C_{21}^2C_{11} & = & 2iL_{P2} L_{\Delta} N_{\Delta} J \label{c32eq4} \\
-i{\rm Tr} \; C_{12}^2C_{22} & = & 4i N_{\Sigma} L_1 L_{\Delta} N_{\Delta} J \label{c32eq5} \\
-i{\rm Tr} \; C_{21}^2C_{22} & = & 4i L_{\Sigma} N_1 L_{\Delta} N_{\Delta} J \label{c32eq6} \\
-i{\rm Tr} \; C_{22}^2C_{12} & = & 2iN_{\Sigma} L_{P2} L_{\Delta} N_{\Delta} J \label{c32eq7} \\
-i{\rm Tr} \; C_{22}^2C_{21} & = & 2iL_{\Sigma} N_{P2} L_{\Delta} N_{\Delta} J \label{c32eq8} \\
-i{\rm Tr} \; C_{11}^2C_{22} & = & 2i(T_{11}+L_1N_1) L_{\Delta} N_{\Delta} J \label{c32eq9} \\
-i{\rm Tr} \; C_{12}^2C_{21} & = & 2i(-T_{12}+L_1N_1^2) L_{\Delta} N_{\Delta} J \label{c32eq10} \\
-i{\rm Tr} \; C_{21}^2C_{12} & = & 2i(-T_{21}+L_1^2N_1)L_{\Delta} N_{\Delta} J \label{c32eq11} \\
-i{\rm Tr} \; C_{22}^2C_{11} & = & 2i(T_{22}+L_1^2N_1^2 -(L_1^2+L_2)(N_1^2+N_2)/4)
                                                      L_{\Delta} N_{\Delta} J \label{c32eq12} 
\end{eqnarray}
The first eight of these (Eqs.~\ref{c32eq1}-\ref{c32eq8}) have no dependence 
on mixing angles except through $J$ as before,
and together with Eq.~\ref{c3eq1} above would in fact be more than sufficient
to fix all the masses in terms of cubic commutators alone,
if that was desirable.
The last four traces however (Eqs.~\ref{c32eq9}-\ref{c32eq12}) 
having no triply repeated index in either sector, 
are seen to involve double-moments ($T_{mn}$ \cite{cecj}) of the $P$-matrix:
\begin{eqnarray}
T_{mn} \; = \; {\rm Tr} \; L^m \, N^n \;= \; m_l^T ({\rm diag} \; m_l)^{m-1} \; 
           P  \; ({\rm diag} \; m_{\nu})^{n-1} m_{\nu} \hspace{1.0cm} \label{tmnp} \\
m_l^T = (m_e,m_{\mu}, m_{\tau}) \hspace{2.7cm} m_{\nu}^T = (m_1,m_2,m_3) \hspace{1.7cm}
                                                              \label{tmnv}
\end{eqnarray}
with the moment-weightings here  
defined in terms of the simple mass-vectors $m_l$, $m_{\nu}$. 
Clearly $T_{00}=3$, $T_{01}=N_1$, 
$T_{02}=N_2$ etc.~by unitarity.
The lowest four non-trivial moments
$T_{11}$, $T_{12}$, $T_{21}$., $T_{22}$ appearing
in Eqs.~\ref{c32eq9}-\ref{c32eq12} above, are clearly sufficient 
to determine the $P$-matrix itself, assuming the masses are known.
  
Finally, we consider
the cubic commutator invariants 
with all three commutators different
which are seen to be complex 
(the $\tilde{R}_{m,n}$ are defined just below Eq.~\ref{ptbmr}):
\begin{eqnarray}
-i{\rm Tr} \; C_{11}C_{12}C_{21} & = & \tilde{R}_{11}L_{\Delta} N_{\Delta}
                                    + i(-T_{11}+3L_1N_1)L_{\Delta} N_{\Delta} J 
                                                                   \label{ccad1} \\
-i{\rm Tr} \; C_{12}C_{22}C_{11} & = & \tilde{R}_{12} L_{\Delta} N_{\Delta}
                                    + i(T_{12}+L_1(2N_1^2-N_2))L_{\Delta} N_{\Delta} J \\
-i{\rm Tr} \; C_{21}C_{11}C_{22} & = & \tilde{R}_{21} L_{\Delta} N_{\Delta}
                                    + i(T_{21}+(2L_1^2-L_2)N_1)L_{\Delta} N_{\Delta} J \\
-i{\rm Tr} \; C_{22}C_{21}C_{12} & = & \tilde{R}_{22} L_{\Delta} N_{\Delta} + \nonumber \\
                               &   & i(-T_{22}+(3L_1^2N_1^2-L_1^2N_2-N_1^2L_2+N_2L_2)/2)
                                            L_{\Delta} N_{\Delta} J.  \label{ccad4} 
\end{eqnarray}
Readily constructed from the $P$ and $K$ matrices,
the $R$-matrix is real and given by
\footnote{
The $R_{l \nu}$ are also expressible in terms
of differences of ``hexaplaquettes'' 
(which are mixing invariants
comprising products of six mixing matrix elements,
choosing two elements (one complex-conjugated) from each row and column,
as noted by Jarlskog and Kleppe~\cite{klep}). We have:
\begin{eqnarray}
\Omega_{l \, \nu}^{\pm} & := &  (U_{l-1 \, \nu} U^*_{l-1 \, \nu+1} 
                                U_{l \,\nu+1}U^*_{l \, \nu-1}
                                    U_{l+1 \, \nu-1}U^*_{l+1 \, \nu}) \nonumber \\
               &   &  \hspace{2.0cm} -(U_{l-1 \, \nu-1} U^*_{l-1 \, \nu} 
                                U_{l \,\nu+1}U^*_{l \, \nu-1}
                                         U_{l+1 \, \nu}U^*_{l+1 \, \nu+1})^{**/*} \label{hexdif} \\
               & := &
             -( (P_{l-1 \, \nu+1}\Pi_{l \, \nu+1}+P_{l+1 \, \nu-1}\Pi^*_{l \, \nu-1})
              - (P_{l-1 \, \nu-1}\Pi_{l-1 \,\nu} +P_{l+1 \, \nu+1}\Pi^*_{l+1 \, \nu})^{**/*}) \\
               & := & -R_{l \, \nu} 
              -i( (P_{l-1 \, \nu+1}-P_{l+1 \, \nu-1})
                              \mp (P_{l-1 \, \nu-1}-P_{l+1 \, \nu+1}))J
\end{eqnarray}
where the sign ambiguity correlates 
with the absence or presence of a relative complex conjugation
in the difference, indicated in Eq.~\ref{hexdif} 
by the overall $**/*$ appearing on the second product.
Hexaplaquettes feature for example in calculations 
of the electric dipole moment of the neutron \cite{edm}.}:
\begin{equation}
R_{l \, \nu} = (P_{l-1 \, \nu-1}K_{l-1 \, \nu} - P_{l-1 \, \nu+1}K_{l \, \nu+1}
            + P_{l+1 \, \nu+1}K_{l+1 \, \nu} - P_{l+1 \, \nu-1}K_{l \, \nu-1}).
\end{equation}
For example, the $R$-matrix for tribimaximal mixing is given by:
\begin{eqnarray}
P= \left( \matrix{ 2/3 & 1/3 & 0 \cr
                   1/6 & 1/3 & 1/2 \cr
                   1/6 & 1/3 & 1/2 } \right)
\hspace{8mm} \Rightarrow \hspace{8mm}
R= \left( \matrix{ 0 & 0 & 0 \cr
                    -1/18 & 1/18 & 0 \cr
                    1/18 & -1/18 & 0 } \right). \label{ptbmr}
\end{eqnarray}
In Eqs.~\ref{ccad1}-\ref{ccad4} (RHS),
the real parts
are proportional to the $R$-matrix moments,
which are written simply $\tilde{R}_{m n}$
($\tilde{R}_{mn} \, := \, m_l^T ({\rm diag} \; m_l)^{m-1} \, 
           R  \, ({\rm diag} \; m_{\nu})^{n-1} m_{\nu}$, cf.~Eq.~\ref{tmnp}).
The associated imaginary parts
are proportional to $J$, 
but depend linearly also on the $T_{mn}$. 
Note that
the $R$-matrix for trimaximal mixing is simply the null matrix,
identical in fact to the $R$-matrix for no-mixing at all,
whereby it is already clear that we cannot 
unambiguously recover the $P$ and $K$ matrices 
starting from the $R$-matrix in general.


Considering cubic and quadratic traces (Section~3) together,
we clearly have here a functionally (over-)complete
set of flavour-symmetric 
Jarlskog-invariant mixing variables,
simply related to the $P$ and $K$ matrices (Sections~1-2). 
Assuming that 
flavour symmetry and Jarlskog invariance 
are always to be respected \cite{binv},
we may anticipate that our present analysis 
will prove pertinent 
to formulating future theories of flavour. 
\\

\noindent {\bf Acknowledgements }  
\vspace{1mm}

\noindent 
This work was supported by the UK 
Particle Physics and Astronomy Research Council (PPARC).
PFH and TJW acknowledge the hospitality
of the Centre for Fundamental Physics (CfFP)
at CCLRC Rutherford Appleton Laboratory.
We are grateful to R.~Kershaw for
independently cross-checking many of the formulae presented herein.
\\

\noindent {\bf Appendix A} 
\vspace{2mm}


\noindent
For the $P$-matrix we know that 
we need only the four elements of any plaquette 
to obtain all the elements, since any row or column 
of the $P$-matrix sums to unity.
Analogously, for the $K$-matrix we can calculate the full matrix
(at least up to a twofold ambiguity, see below)
in terms of the four elements of any $K$-plaquette.
We need the sum ($S_{l\nu}$) of the elements of the $K$-plaquette,
the product of all four elements ($F_{l\nu}$), 
and the sum of the products taken in threes ($W_{l\nu}$)
and also the determinant ($D_{l\nu}$):
\begin{eqnarray}
S_{l\nu}=K_{l+1 \; \nu+1}+K_{l-1 \; \nu+1}+K_{l+1 \; \nu-1}+K_{l-1 \; \nu-1} \\
F_{l\nu}=K_{l+1 \; \nu+1} K_{l-1 \; \nu+1} K_{l+1 \; \nu-1} K_{l-1 \; \nu-1} \\
W_{l\nu}=F_{l\nu} \; ( \, 1/K_{l+1 \; \nu+1}+1/K_{l-1 \; \nu+1}+1/K_{l+1 \; \nu-1}+1/K_{l-1 \; \nu-1} \,) \\
D_{l\nu}=K_{l+1 \; \nu +1}K_{l-1\; \nu-1}- K_{l-1 \; \nu +1}K_{l+1 \; \nu-1}
\end{eqnarray}
where the formula for $W_{l\nu}$ has been written here in terms of $F_{l\nu}$ only for brevity 
(ie.\ in the case of any $K$-plaquette elements being zero,
it is to be understood that those denominators are to be cancelled 
with the corresponding factor in the numerator before 
proceeding with the substitution.)  
The square of the Jarlskog invariant $J$ is given by:
\begin{equation}
J^2=( \; \frac{W_{l\nu}}{S_{l\nu}^2} \; + \; \frac{D_{l\nu}^2}{S_{l\nu}^3} \; )
                ( \; \frac{1}{2} \; \pm \; \sqrt{\frac{1}{4}-S_{l\nu}} \; ) 
                                                 \; - \; \frac{D_{l\nu}^2}{S_{l\nu}^2}. \label{j2kp}
\end{equation}
The $K$-elements not in the $K$-plaquette 
(with the exception of the $K_{l \; \nu}$ element itself)
can then normally be obtained using, eg.\ :
\begin{eqnarray}
K_{l + 1 \; \nu}=(J^2-K_{l + 1 \; \nu+1} K_{l + 1 \; \nu-1})/
                                        (K_{l + 1 \; \nu+1}+K_{l + 1 \; \nu-1})
\end{eqnarray}
with analogous formulae for $K_{l -1 \; \nu}$, $K_{l \; \nu + 1}$ etc.
The $K_{l \nu}$ element itself is given by:
\begin{eqnarray}
K_{l \; \nu} = -(J^4-(A_{l\nu}+P_{l\nu})J^2+F_{l\nu})/(S_{l\nu} \, J^2 -W_{l\nu}).
\end{eqnarray}

As an example of these procedures 
we shall first consider the $K$-matrix for trimaximal mixing Eq.~\ref{ktri}
(since the tribimaximal case requires a little more more care).
Suppose that only the $K_{\tau 3}$ 
plaquette is known, ie. we have 
$K_{e1}=1/18$, $K_{e2}=1/18$, $K_{\mu 2}=1/18$, $K_{\mu 1}=1/18$,
but no a priori knowledge of 
$K_{e 3}$, $K_{\mu 3}$, $K_{\tau 1}$, $K_{\tau 2}$ or $K_{\tau 3}$.
We wish to reconstruct the full trimaximal $K$ matrix: 
\begin{eqnarray}
K= \left( \matrix{ 1/18 & 1/18 & * \cr
                   1/18 & 1/18 & * \cr
                      * & * & * } \right)
\hspace{8mm} \Rightarrow \hspace{8mm}
K= \left( \matrix{ 1/18 & 1/18 & 1/18 \cr
                   1/18 & 1/18 & 1/18 \cr
                   1/18 & 1/18 & 1/18 } \right) \label{ptok}
\end{eqnarray}
From the $K_{\tau 3}$ plaquette (Eq.~\ref{ptok} LHS) we have:
\begin{equation} 
S_{\tau 3}= 4/18 \hspace{7mm} F_{\tau 3}=1/18^4 \hspace{7mm}
W_{\tau 3}= 4/18^3 \hspace{7mm} D_{\tau 3} = 0
\end{equation}
We may then calculate $J^2$ from Eq.~\ref{j2kp} as follows:
\begin{eqnarray}
J^2 & =  & (1/(4 \times 18)+0)(1/2 \pm \sqrt{1/4-4/18})-0 \\
     & = &  \frac{1}{4 \times 18} \times (\frac{1}{2} \pm \frac{1}{6})
       = \frac{1}{4 \times 27} \hspace{3mm} {\rm or} 
         \hspace {3mm} \frac{1}{8 \times 27} 
\end{eqnarray}
Taking $J^2=1/(4 \times 27)$ corresponding 
to maximal $CP$ violation ($J=J_{\rm max}=\pm 1/6\sqrt{3}$),
ie.\ corresponding to trimaximal mixing, we have, eg.\ 
\begin{equation}
K_{e 3}= \frac{1/(4 \times 27)-1/18 \times 1/18}{1/18+1/18} 
          =\frac{1/(2 \times 81)}{2/18}=1/18
\end{equation}
and 
\begin{eqnarray}
K_{\tau 3} & = & -\frac{1/(4 \times 27)^2-(4/18^2+2/18^2))/(4 \times 27)+1/18^4}
           {4/18 \times 1/(4 \times 27)-4/18^3} \\
           & = & -\frac{(9-18+1)/(2^4 \times 3^8)}{(3-1)/(2 \times 3^6)}=\frac{1}{18}
\end{eqnarray}
which leads to Eq.~\ref{ptok}.
Taking instead the alternative solution 
$J^2=1/(8 \times 27)$ (with intermediate $CP$ violation $J=J_{\rm max}/\sqrt{2}=\pm 1/6\sqrt{6}$) 
leads to the $K$ matrix (RHS):
\begin{eqnarray}
P= \left( \matrix{ 5/12 & 5/12 & 1/6 \cr
                   5/12 & 5/12 & 1/6 \cr
                   1/6 & 1/6 & 2/3 } \right)
\hspace{8mm} \Leftrightarrow \hspace{8mm}
K= \left( \matrix{ 1/18 & 1/18 & 1/72 \cr
                   1/18 & 1/18 & 1/72 \cr
                   1/72 & 1/72 & 23/144 } \right)
\end{eqnarray}
corresponding to the uninteresting $P$ matrix also displayed (LHS).
Thus the $K$ matrix is determined from a $K$ plaquette here,
only up to a two-fold ambiguity.

In case of zeros in the denominators above,
we must use less succinct but slightly more general formulae 
(as in the definition of $W_{l\nu}$).
The element $K_{l+1 \; \nu}$ then becomes:
\begin{eqnarray}
K_{l+1 \; \nu}=\frac{B_{l+1 \; \nu}}{S_{l\nu}^3}(\frac{1}{2} \pm \sqrt{\frac{1}{4}-S_{l\nu}}) 
         -\frac{K_{l + 1 \; \nu+1}K_{l + 1 \; \nu-1}}{S_{l\nu}} \nonumber \hspace{6.2cm} \\
  -\frac{K_{l + 1 \; \nu+1}K_{l + 1 \; \nu-1}
                (K_{l - 1 \; \nu+1}+K_{l - 1 \; \nu-1})
+K_{l + 1 \; \nu+1}K_{l - 1 \; \nu-1}^2
                     +K_{l + 1 \; \nu-1}K_{l - 1 \; \nu+1}^2}{S_{l\nu}^2} \hspace{2mm}
\end{eqnarray}
where $B_{l+1 \; \nu}$ is the product of three sums of adjacent pairs of elements as follows:
\begin{equation}
B_{l+1 \; \nu}=(K_{l + 1 \; \nu+1}+K_{l - 1 \; \nu+1})
                             (K_{l + 1 \; \nu-1}+K_{l - 1 \; \nu-1})
                                   (K_{l - 1 \; \nu+1}+K_{l - 1 \; \nu-1})
\end{equation}
ie. excluding the pair in line with the element required.
The $K_{l \nu}$ element is given by:
\begin{equation}
K_{l \nu}=\frac{2S_{l\nu}G_{l\nu}+H_{l\nu}(1\pm\sqrt{1-4S_{l\nu}})}
                                  {S_{l\nu}^4(2S_{l\nu}-(1\pm\sqrt{1-4S_{l\nu}}))}
\end{equation}
where $G_{l\nu}$ and $H_{\l\nu}$ are defined by:
\begin{eqnarray}
G_{l\nu} & := & 8S_{l\nu}F_{l\nu}+S_{l\nu}P_{l\nu}(S_{l\nu}^2-2M_{l\nu})
                              +3D_{l\nu}^2S_{l\nu}-D_{l\nu}^2-W_{l\nu}S_{l\nu}(1+S_{l\nu}) \\
H_{l\nu} & := & W_{l\nu}S_{l\nu}-S_{l\nu}^3(A_{l\nu}+M_{l\nu})
                            +(2S_{l\nu}-1)(4F_{l\nu}-M_{l\nu}^2). \hspace{1.6cm}
\end{eqnarray}

Tri-bimaximal mixing is a case in point.
Suppose we wish to reconstruct the tri-bimaximal $K$ matrix,
starting only from its $K_{\tau 3}$ plaquette:
\begin{eqnarray}
K= \left( \matrix{ 1/6 & 1/12 & * \cr
                     0 & 0 & * \cr
                      * & * & * } \right)
\hspace{8mm} \Rightarrow \hspace{8mm}
K= \left( \matrix{ 1/6 & 1/12 & -1/18 \cr
                    0 & 0 & 1/9 \cr
                    0 & 0 & 1/9 } \right) \label{pbok}
\end{eqnarray}
in which case we have:
\begin{equation} 
A_{\tau 3}=1/72 
\hspace{10mm}
S_{\tau 3}=1/4 
\hspace{10mm} M_{\tau 3}=F_{\tau 3}=
W_{\tau 3}= D_{\tau 3} = 0.
\end{equation}
From Eq.~\ref{j2kp} the $CP$ violation is zero ($J^2=0$)
and, eg.\ the $K_{e3}$ element is given by:
\begin{equation}
K_{e3}=-\frac{1/6 \times 1/12}{1/4}=-\frac{4}{6 \times 12}=-\frac{1}{18}.
\end{equation} 
We have then also:
\begin{equation} 
G_{\tau 3}=0 \hspace{15mm}
H_{\tau 3}=-1/4^3 \times 1/72.
\end{equation}
so that the $K_{\tau 3}$ element is given by:
\begin{equation}
K_{e3}=-\frac{-1/4^3 \times 1/72}{1/4^4 \times (-1/2)}=\frac{4 \times 2}{72}=\frac{1}{9}
\end{equation} 
with no ambiguities arising
(clearly no ambiguitities arise whenever $S_{l \nu}=1/4$). \\

%

\end{document}